\documentclass[11pt,fleqn]{article}

\usepackage{amsmath}
\usepackage{graphicx}

\usepackage{amsfonts,amssymb,cite}
\usepackage{epsf,graphicx}

\def\noi{\noindent}
\def\jnumber#1#2{\thispagestyle{empty} \noi\unitlength=1mm
    	\begin{picture}(178,10)
            \put(177,15){\llap{\large\it Grav. Cosmol. No.\,#1, #2}}
                    \end{picture}}

\newcommand{\Title}[1]{\noi {{\Large\bf #1}}\\[1ex]}

\newcommand{\Author}[2]{\noi{\bf #1}\\[2ex]\noi{\normalsize\it #2}\\}

\newcommand{\Abstract}[1]{\vskip 2mm \begin{center}
        \parbox{16.4cm}{\small\noi #1} \end{center}\medskip}

\newcommand{\foom}[1]{\protect\footnotemark[#1]}

\def\email#1#2{\footnotetext[#1]{e-mail: #2}\addtocounter{footnote}{1}}

\def\nqq{\hspace*{-2em}}

\def\cm{\hspace*{1cm}}

\def\Jl#1#2{#1 {\bf #2},\ }

\def\ApJ#1 {\Jl{Astroph. J.}{#1}}
\def\CQG#1 {\Jl{Class. Quantum Grav.}{#1}}
\def\DAN#1 {\Jl{Dokl. AN SSSR}{#1}}
\def\GC#1 {\Jl{Grav. Cosmol.}{#1}}
\def\GRG#1 {\Jl{Gen. Rel. Grav.}{#1}}
\def\JETF#1 {\Jl{Zh. Eksp. Teor. Fiz.}{#1}}
\def\JETP#1 {\Jl{Sov. Phys. JETP}{#1}}
\def\JHEP#1 {\Jl{JHEP}{#1}}
\def\JMP#1 {\Jl{J. Math. Phys.}{#1}}
\def\NPB#1 {\Jl{Nucl. Phys. B}{#1}}
\def\NP#1 {\Jl{Nucl. Phys.}{#1}}
\def\PLA#1 {\Jl{Phys. Lett. A}{#1}}
\def\PLB#1 {\Jl{Phys. Lett. B}{#1}}
\def\PRD#1 {\Jl{Phys. Rev. D}{#1}}
\def\PRL#1 {\Jl{Phys. Rev. Lett.}{#1}}

\def\lal{&&\nqq {}}
\def\eq{Eq.\,}
\def\eqs{Eqs.\,}
\def\beq{\begin{equation}}
\def\eeq{\end{equation}}
\def\bear{\begin{eqnarray}}
\def\bearr{\begin{eqnarray} \lal}
\def\ear{\end{eqnarray}}
\def\earn{\nonumber \end{eqnarray}}

\def\nnn{\nonumber\\ \lal }

\def\yy{\\[5pt] {}}
\def\yyy{\\[5pt] \lal }

\def\const{{\rm const}}
\def\eps{\varepsilon}

\newcommand{\aver}[1]{\langle \, #1 \, \rangle \mathstrut}

\begin{document}
\twocolumn[
\jnumber{3}{2013}
\vspace*{-8mm}

\Title{Dark energy from instantons}

\Author{Leonid Marochnik\foom 1}
{Physics Department, East-West Space Science Center,
\it University of Maryland, College Park, MD 20742 USA}



\Abstract
  {We show that in imaginary time quantum metric fluctuations of empty space
   form a self-consistent de Sitter gravitational instanton that can be
   thought of as describing tunneling from ``nothing" into de Sitter
   space of real time (no cosmological constant or scalar fields are needed).
   For the first time, this mechanism is activated to give birth to a flat
   inflationary Universe. For the second time, it is turned on to complete
   the cosmological evolution after the energy density of matter drops below
   the threshold (the energy density of instantons). A cosmological expansion
   with dark energy takes over after the scale factor exceeds this threshold,
   which marks the birth of dark energy at a redshift $1+z\approx 1.3$ and
   provides a possible solution to the ``coincidence problem''. The number
   of gravitons which tunneled into the Universe must be of the order of
   $10^{122}$ to create the observed value of the Hubble constant. This
   number has nothing to do with vacuum energy, which is a possible solution
   to the ``old cosmological constant problem''. The emptying Universe should
   possibly complete its evolution by tunneling back to ``nothing''.  After
   that, the entire scenario is repeated, and it can happen endlessly.
   }

\begin{flushright}
\parbox{13cm}{\small\noi
    ``One might think this means that imaginary numbers are just a
    mathematical game having nothing to do with the real world. From the
    viewpoint of positivist philosophy, however, one cannot determine
    what is real. All one can do is find which mathematical models
    describe the universe we live in.''\yy
    \cm Stephen Hawking,
    {\it The Universe in Nutshell\/}, Bantam Books, 2001, p.\,59.
    }
\end{flushright}

\bigskip

] 
\email 1 {lmarochnik@gmail.com}

{ 
\def\hpsi {{\hat \psi}{}}
\def\bpsi {{\bar \psi}{}}

\section{Introduction}

  The cosmological acceleration (dark energy (DE) effect) was discovered by
  observations of supernova SNIa by Riess et al. [1] and Perlmutter et al.
  [2]. Since then a number of hypotheses were advanced to explain this
  phenomenon (for references to original work, see, e.g., [3-7] and
  references therein; for a recent development see, e.g., [8]). The
  constraint on the equation-of-state parameter of DE $w = p_{\rm
  de}/\eps_{\rm de}$ is $w=-1.08\pm 0.1$ [9]. This equation of state
  corresponds to the de Sitter solution for the empty isotropic and
  homogeneous Universe with a nonzero positive cosmological constant
  $\Lambda$. Although the $\Lambda$ term is consistent with the
  observational value of $w\approx -1$, there are well-known problems with
  that. The first one is the so-called ``old cosmological constant
  problem'':  Why is $\Lambda$ measured from observations of the order of
  $10^{-122}$ of vacuum energy density? The second one is the ``coincidence
  problem'': Why is the acceleration happening during the contemporary epoch
  of matter domination? As a matter of fact, the observational constraint
  $w_{\rm de}\approx -1$ tells us only that the DE equation of state
  is close to $p=-\eps$, and this does not mean that it is necessarily due
  to $\Lambda\neq 0$. Such an equation of state can exist for other reasons
  (a well-known example is a scalar field). In the present work, we assume
  that the DE is of instanton origin, and such an assumption seems to be
  able to resolve the issues mentioned above.

  In general, instantons are Euclidean solutions that mediate tunneling
  between two vacua (see, e.g., [10] and references therein). The idea of the
  present work to employ gravitational instantons of a certain type to tunnel
  ``something'' to the Lorentzian space of real time is not new. Basic works
  on gravitational instantons are collected in [11]. In application to
  quantum cosmology, the basic idea was to explain the origin of an
  inflationary Universe of Lorentzian signature by tunneling from a
  Riemannian space of Euclidean signature or from ``nothing''. There is the
  well-known work by Tryon [12] who was probably the first to propose that
  our Universe could be a vacuum fluctuation, that ``... our Universe did
  indeed appear from nowhere'' and to mention that ``... such an event need
  not have violated any of the conventional laws of physics''; Zeldovich
  [13] who discussed a quantum creation of the Universe; Atkatz and Pagel
  [14] who proposed that ``... the Universe arises as a result of
  quantum-mechanical barrier penetration''; the ``no boundary proposal'' by
  Hartle and Hawking [15] and a birth of the inflationary Universe by
  tunneling from ``nothing'' by Vilenkin [16] and Grishchuk and Zeldovich
  [17]. In the framework of quantum gravity, tunneling from a Riemannian
  space of Euclidean signature was considered by Gibbons and Hartle [18]
  and others.

  In this paper, we show that this old idea can get a ``second wind'' due to
  a mechanism of tunneling based on quantum metric fluctuations, which is
  able to address both the birth of DE and inflationary Universe. As well as
  in [16], by ``nothing'' we mean a state with no classical space-time.
  Vilenkin's [16] proposal was based on the Hawking-Moss instanton [19] which
  gives birth to a closed inflationary universe for some models of scalar
  field.\footnote
    {An attempt to create a flat inflationary universe by a
    gravitational instanton formed by a scalar field was made in [26]
    but met criticism (see [27, 28]; see also [29, 30]).}
  We believe that the advantage of the mechanism proposed here lies in the
  fact that it is quite universal in the sense that it does not depend on
  the model of a scalar field but is based on natural quantum fluctuations
  of the space metric. It is also able to give birth to both a flat
  inflationary Universe to start its cosmological evolution and to DE in the
  contemporary epoch of aging emptying Universe to finish its cosmological
  evolution. This fact permits us to interpret both phenomena by a single
  mechanism (for similarities between primordial DE driving inflation and
  present DE see [20]). The Einstein equations do not fix the signature,
  which means that a signature change can also be considered in the
  framework of classical general relativity ([21--23] and others; see also
  [24] for a complete list of references). This approach can be thought of
  as a classical idealization of tunneling solutions [25].

  Technically, in this paper, we find self-consistent solutions to the
  equations of quantum gravity in the one-loop approximation in imaginary
  time and then analytically continue these to the Lorentzian space of real
  time. Assuming that these solutions do exist in the space of real time,
  we show that this procedure is capable of providing a plausible
  interpretation for both the birth of a flat inflationary universe and the
  DE.  In a sense, this is one of possible concrete realizations of
  Hawking's idea on the reality of imaginary time. We can only partly
  overcome the present lack of a consistent quantum theory of gravity by
  using the one-loop approximation in which it is finite and mathematically
  consistent (see [31, 32]). The one-loop approximation of quantum gravity
  is believed to be applicable to the modern Universe because of its
  remoteness from the Planck epoch. As was shown in [31, 32], the de Sitter
  gravitational instanton is one of three exact solutions to the exact
  self-consistent equations of one-loop quantum gravity that were obtained
  by using the BBGKY chain approach. In this paper, we obtain the same de
  Sitter exact solution directly from the original equations of one-loop
  quantum gravity (with no use of the BBGKY chain). This approach creates a
  window of opportunity for the new ``tunneling interpretation'' of this
  solution, which does not require``ghost materialization'' in real time that
  was the case in [32].  Another goal of this work is to show that the
  tunneling of DE to the real-time Universe has favorable conditions
  precisely in the matter-dominated epoch (Section 4).

  In Section 2, we present exact self-consistent equations of quantum gravity
  in the one-loop approximation in real time. We show that in real time
  quantum metric fluctuations are unable to form a self-consistent de Sitter
  solution, but they can do that if the time variable has at least an
  infinitesimally small imaginary part. In Section 3, we show that in
  imaginary time (Euclidean space) quantum metric fluctuations form a
  self-consistent de Sitter instanton that can be thought of as describing
  tunneling to the Lorentzian space of real time from ``nothing''. The
  topological non-equivalence between manifolds plays a role of a classically
  impenetrable barrier, quantum tunneling across which can create a flat
  inflationary Universe. In Section 4, we show that in the presence of
  matter, such tunneling is able to give birth to DE only after the density
  of matter drops below a critical level, and the Universe will become
  quite empty again. The existence of such a threshold is a possible
  solution to the ``coincidence problem''. In Section 5, we outline
  a scenario of cosmological evolution based on the proposed mechanism.


\section{Quantum metric fluctuations in real time}

  In this work, for curvatures much smaller than the Planck curvature, quantum
  cosmology is represented as a theory of gravitons in macroscopic
  space-time with a self-consistent geometry. The quantum state of gravitons
  is determined by their interaction with a macroscopic field, and the
  macroscopic (background) geometry, in turn, depends on the state of
  gravitons. The background metric and the graviton operator appearing in the
  self--consistent theory are extracted from the unified gravitational field,
  which initially satisfies the exact equations of quantum gravity. The
  classical component of the unified field is a function of coordinates and
  time by definition. The quantum component of the same unified field is
  described by a tensor operator function, which also depends on coordinates
  and time. Under such formulation of the problem, the original exact
  equations should be the operator equations of quantum theory of gravity in
  the Heisenberg representation. The rigorous mathematical derivation of
  these equations and their relation to existing references can be found in
  [33]. For the first time, these equations (and exact solutions) were
  given in [31]. Referring the reader for details to these works (see also
  [32], sections II, III and XII), note that an inherent part of these
  equations is the unavoidable appearance of the auxiliary ghost fields
  introduced by Feynman [34] and known as Faddeev-Popov ghosts [35]. It is
  precisely the appearance of ghosts that ensures the one-loop finiteness of
  quantum gravity, making the theory mathematically consistent ([31], [32]
  Sections III.D and III.E). In the self-consistent theory of gravitons, the
  background metric is described by regular vacuum Einstein equations [31]
\bearr            \label{mar-eq1}
    R^k_i-\frac{1}{2}\delta^k_iR
    =\kappa \big( \aver{ \Psi_{g}|\hat{T}^k_{i(\rm grav)}|\Psi_{g} }
\nnn \cm
    + \aver{\Psi_{\rm gh}|\hat{T}^k_{i(\rm ghost)}|\Psi_{\rm gh}} \big).
\ear

  Here $\Psi_\mathrm{g}$, $\Psi_\mathrm{\rm gh}$ are quantum state vectors of
  gravitons and ghosts, respectively. The explicit form of the
  energy-momentum tensors of gravitons $\hat{T}^k_{i(\rm grav)}$ and
  $\hat{T}^k_{i(\rm ghost)}$ is presented in [32], Section II.F). The second
  term on the right-hand side of (\ref{mar-eq1}) comes from Faddeev-Popov
  ghosts and provides one-loop finiteness of quantum gravity. We consider
  the self-consistent model of the Universe which is flat, isotropic and
  homogeneous on the average (the FLRW metric). The calculations presented
  here were done in the class of synchronous gauges (that automatically
  provide one-loop finiteness of observables). From (\ref{mar-eq1}) follow
  equations for the energy density $\eps_\mathrm{g}$ and pressure
  $p_\mathrm{g}$ ([31] and [32], Section III.B). Heisenberg's operator
  equations for Fourier components of the transverse 3-tensor graviton
  field, Grassmann's ghost field and canonical commutation relations for
  gravitons and anti-commutation relations for ghosts are also presented in
  these papers. These equations form a self-consistent set of equations of
  one-loop quantum gravity for gravitons, ghosts and the FLRW background
  (\eqs (2)--(4) of [31]. Another form of the same equations is presented
  in [32] (\eqs (III.30-III.34)). In this paper, we use the latter, in
  which it is convenient to use the conformal time $\eta = \int{dt/a}$ and
  to pass on from summing to integration by the transformation
\[
    \sum_k...\to\int{d^3k/(2\pi)^3...}
        =\int_0^{\infty}{k^2dk/2\pi^2...}.
\]
  These operations lead to the following set of equations:
\bearr                    \label{mar-eq2}
    3\frac{a'^2}{a^4 }=\kappa \eps_{g} =\frac{1}{16\pi^2}\int_0^\infty
    {\frac{ k^2}{ a^2}} dk
\nnn \cm    \times
    \Big(\sum_\sigma \aver{\Psi_g |\hpsi'_{{\bf k}\sigma}{}^{+}
        \hpsi'_{{\bf k}\sigma}  +k^2\hpsi_{{\bf k}\sigma}^{+}
        \hpsi_{{\bf k}\sigma } | \Psi_g }
\nnn \cm
    -2 \aver{\Psi_{\rm gh} | \bar{\theta}{}'_{\bf k}\theta'_{\bf k}
    + k^2 \bar{\theta}_{\bf k} \theta_{\bf k}| \Psi_{\rm gh} } \Big),
\yyy                    \label{mar-eq3}
    2\frac{a''}{a^3}-\frac{a'{}^2}{a^4}= -\kappa p_g
    = -\frac{1}{16\pi^2}\int_0^\infty \frac{k^2}{a^2} dk
\nnn \cm \times
    \Big( \sum_\sigma \aver{\Psi_g |\hpsi_{{\bf k}\sigma}^{'+}
    \hpsi'_{{\bf k}\sigma} -\frac{k^2}{3} \hpsi_{{\bf k}\sigma}^{+}
    \hpsi_{{\bf k}\sigma} | \Psi_g }
\nnn \cm
    -2 \aver{\Psi_{\rm gh} | \bar{\theta}{}'_{\bf k}
    \theta'_{\bf k} - \frac{k^2}{3}\bar{\theta}_{\bf k}
        \theta_{\bf k}| \Psi_{\rm gh} } \Big).
\ear
  And for fluctuations
\bearr                                            \label{mar-eq4}
    \hat{\phi}''_{{\bf k},\sigma}
     + \big(k^2 {-} a''/a\big) \hat{\phi}_{{\bf k},\sigma} = 0,
   \quad
    \hat{\psi}_{k\sigma} = (1/a) \hat{\phi}_{k\sigma}
\yyy                    \label{mar-eq5}
   \hat{\vartheta}''_{\bf k}
    +\big (k^2 - a''/a)\hat{\vartheta}_{\bf k}=0,
   \qquad
    \hat{\theta}_{\bf k} = (1/a)\hat\vartheta_{\bf k}
\ear
  Here $\sigma$ is the polarization index; $a(\eta)$ is the FLRW scale
  factor; $\kappa = 8\pi G$; the superscript ``+'' denotes complex
  conjugation, and dots are time derivatives. Primes denote derivatives in
  the conformal time $\eta$.  The de Sitter expansion is
\beq                                                     \label{mar-eq6}
    a_s=-(H\eta)^{-1}
\eeq
  In Section 3, we will need exact solutions to (\ref{mar-eq4}),
  (\ref{mar-eq5}) in the de Sitter background (\ref{mar-eq6}). They read
  (see [31])
\bearr                        \label{mar-eq7}
    \hpsi_{{\bf k}\sigma}=\frac{1}{a_s}\sqrt{\frac{2\kappa\hbar}{k}}
    \left[\hat{c}_{{\bf k}\sigma}f(x)
        +\hat{c}^+_{-{\bf k}-\sigma}f^+(x)\right],
\nnn
    \hat{\theta}_{\bf k}=\frac{1}{a_s}\sqrt{\frac{2\kappa\hbar}{k}}
    \left[\hat{\alpha}_{\bf k}f(x)+\hat{\bar{\beta}}_{-{\bf k}}f^+(x)\right],
\yyy            \label{mar-eq8}
    f(x) = (1- i/x) e^{-ix}, \qquad     x = k\eta.
\ear
  Quantum metric fluctuations (\ref{mar-eq7}), (\ref{mar-eq8}) are unable to
  form a self-consistent de Sitter solution to \eqs (\ref{mar-eq7})%
  --(\ref{mar-eq8}) in real time because of incomputability of the
  integrals $\int_0^{\infty}{x^n e^{\pm 2 i x}dx}$  arising in the
  right-hand side of (\ref{mar-eq2}), (\ref{mar-eq3}). However, they can do
  that if these incomputable integrals are redefined as [31]
\[
    \int_0^{\infty} x^n e^{\pm 2 i x} dx = \lim\limits_{\delta\to 0}
    \int_0^{\infty} x^n e^{\pm 2 i x-\delta x} dx.
\]
  Such a re-definition implies that the variable $x = k\eta$ must have at
  least an infinitesimally small imaginary part to be able to form a
  self-consistent de Sitter solution. In the next section, we show that the
  self-consistent de Sitter solution does exist in imaginary time, i.e., in
  Riemannian space of Euclidean signature from which it can be
   analytically continued to the Lorentzian space of real time.


\section{Quantum metric fluctuations in imaginary time}

  The transition to imaginary time in \eqs (\ref{mar-eq2})-(\ref{mar-eq5})
  is carried out by replacing the variables:
\beq                                         \label{mar-eq9}
    t = -i\tau,  \qquad   \eta=-i\upsilon
\eeq
  The transition (\ref{mar-eq9}) transforms (\ref{mar-eq2}-\ref{mar-eq5}) to
  the following set of equations:
\bearr                        \label{mar-eq10}
    -3\frac{a'^2}{a^4}=\frac{1}{16\pi^2} \int_0^{\infty}\frac{k^2}{a^2}dk
\nnn \cm \times
    \biggl(\sum_{\sigma} \aver{\Psi_g|-\hpsi'^+_{k\sigma}\hpsi'_{k\sigma}
    +k^2\hpsi^+_{k\sigma}\hpsi_{k\sigma}|\Psi_g }
\nnn \cm
    - 2 \aver{\Psi_{\rm gh}| -\bar{\theta}'_k\theta'_k
    + k^2{\bar\theta}_k \theta_k|\Psi_{\rm gh} } \biggr)
\ear
\bearr                    \label{mar-eq11}
    \hat{\phi''}_{{\bf k},\sigma}
    -(k^2 + a''/a)\hat{\phi}_{{\bf k}, \sigma}=0,
   \quad
    \hpsi_{k\sigma} = \frac{1}{a} \hat{\phi}_{k\sigma};
\yyy                    \label{mar-eq12}
   \hat{\vartheta''}_{\bf k} -(k^2 + a''/a)\hat\vartheta_{\bf k}=0,
   \qquad
    \hat{\theta}_k = (1/a) \hat\vartheta_{\bar{k}}.
\ear
  Primes in this section denote derivatives in the imaginary conformal time
  $\upsilon$. After such a transition, solutions to (10)--(12) over de
  Sitter background read
\bearr                \label{mar-eq13}
    \hpsi_{k\sigma}=\frac{1}{a}\sqrt{\frac{2\kappa \hbar}{k}}
    (\hat{Q}_{\kappa\sigma}g_k + \hat{p}_{k\sigma}h_k),
\nnn
    \hat{\theta}_k=\frac{1}{a}\sqrt{\frac{2\kappa \hbar}{k}}
        (\hat{q_\kappa}g_k+\hat{p}_kh_k),
\ear
  where
\bearr   \label{mar-eq14}
    g(\xi) = (1 + 1/\xi)e^{-\xi}, \qquad  h(\xi)=(1-1/\xi)e^{\xi},
\nnn \cm
    \xi=k\upsilon.
\ear
  The requirement of finiteness eliminates the $h$--solution. The operator
  functions (\ref{mar-eq13}) can be named quantum fields of gravitational
  instantons of graviton and ghost type, or for short, graviton-ghost
  instantons. Substitution of (\ref{mar-eq13})-(\ref{mar-eq14}) into the
  right-hand side of (\ref{mar-eq10}) leads to
\bearr      \label{mar-eq15}
    3\frac{a'^2}{a^4}=\frac{\kappa \hbar H_\tau^4}{2\pi^2}
    \int_0^{\infty} N_k [\xi^2-(1+\xi)^2] e^{-2\xi}\xi d\xi,
\yyy        \label{mar-eq16}
    N_k = \sum_{\sigma} \aver{\Psi_g |\hat{Q}^+_{k\sigma}Q_{k\sigma}|
     \Psi_g }
\nnn \cm
     -2 \aver{\Psi_{\rm gh}|\hat{q}_k^+\hat{q}_k|\Psi_{\rm gh}}.
\ear
  Here $H_\tau$ is the Hubble function in imaginary time. The de Sitter
  background in imaginary time is
\beq                              \label{mar-eq17}
    a_s=-(H_{\tau}\upsilon)^{-1}
\eeq

  In the de Sitter background (\ref{mar-eq17}), the left-hand side of
  (\ref{mar-eq15}) must be $3H_\tau^2 = \const$, which means that the
  right-hand side of (\ref{mar-eq15}) cannot be a function of $\upsilon$. In
  turn, this means that only a flat spectrum $N_k = \const$ is able to
  provide constancy of the right-hand side of (\ref{mar-eq15}). To make the
  result more transparent, assume that the spectrum is flat and the numbers
  of instantons of ghost and anti-ghost type are equal to each other, i.e.,
  $\aver{n_{k(g)}} = \aver{n_g}$;  $\aver{n_{k(\rm gh)}}=
  \aver{\bar{n}_{\rm gh}} = \aver{n_{\rm gh}}$. Assume also that typical
  occupation numbers in the ensemble are large, so that squares of modules
  of probability amplitudes are likely to be described by Poisson
  distributions. In such a case, we get a simple physically transparent
  result ([32], section VII):
\beq                                      \label{mar-eq18}
    N_k = 4( \aver{n_g} - \aver{n_{\rm gh}}).
\eeq

  The de Sitter solution (\ref{mar-eq17}) satisfies \eq (\ref{mar-eq15})
  if $H_\tau$ satisfies \eq (\ref{mar-eq19}):
\beq        \label{mar-eq19}
    {H_\tau}^2\left(1+\frac{\kappa\hbar{H_\tau}^2( \aver{n_g}
    - \aver{n_{\rm gh}})} {8\pi^2} \right)=0.
\eeq

  A real solution to (\ref{mar-eq19}) exists if  $\aver{n_{(g)}} -
  \aver{n_{\rm gh}} \leq 0$. It reads
\bearr          \label{mar-eq20}
    {H_\tau}^2=\frac{8\pi^2}{\kappa\hbar(\aver{n_{\rm gh}}-\aver{n_g})}
    \quad \mbox{if} \quad
    {H_\tau}^2 \neq 0,
\nnn
    {H_\tau}^2=0  \quad \mbox{if} \quad
    {H_\tau}^2 \neq \frac{8\pi^2}
        {\kappa\hbar(\aver{ n_{\rm gh}} - \aver{ n_g })}.
\ear

  In real time, ghosts are fictitious particles which appear to compensate
  the spurious effect of vacuum polarization of fictitious fields of
  inertia. In real time, the gravitational effect of gravitons is
  proportional to $(\aver{n_g}-\aver{n_{\rm gh}})\geq 0$ and, figuratively
  speaking, this means that those ghosts cannot be ``materialized" in real
  time. The solution (\ref{mar-eq20}) tells us that in imaginary time
  ghosts are ``materialized'' to form a self-consistent de Sitter solution.
  A remarkable fact is that a passage to the Lorentzian space of real
  time ``de-materializes" the ghosts (see below), so that since the analytic
  continuation is done, the difference (\ref{mar-eq18}) is positive again in
  the Lorentzian space, and ghosts are again fictitious particles as they
  must be.\footnote
    {Ghost ``materialization'' in imaginary time does not affect
    physical quantities in real time. A good example is the creation of
    electron-positron pairs in a constant electric field [16]. Energy
    conservation for the electron is given by the equation
    $m(1- {\dot x}{}^2)^{1/2}-eEx = \const$. To compute the probability
    of pair creation, one needs a transition to imaginary time which
        leads to $m(1 + {\dot x}{}^2)^{1/2}-eEx = \const$. The formal
    appearance of superluminal motion in imaginary time does not affect
    the physical quantities in real time.}
  The next step is to analytically continue the imaginary-time solution
  (\ref{mar-eq17}) to the space of real time. To do so, we analytically
  continue (\ref{mar-eq17}) from the imaginary axis $\upsilon$ to the plane
  of complex conformal time $\varsigma = i\eta + \upsilon$. It reads
\beq                                                      \label{mar-eq21}
    a(\varsigma) = -(H\varsigma)^{-1}.
\eeq
  Here
\beq                    \label{mar-eq22}
    H = \pm\left(\frac{8\pi^2}{\kappa\hbar \aver{n_g}-\aver{n_{\rm gh}}}
        \right)^{1/2} = \pm iH_\tau.
\eeq

  Assuming that the number of gravitons in the Universe is
  $N \approx \aver{n_g} - \aver{n_{\rm gh}}$, we can rewrite $H$ as
\beq                                                   \label{mar-eq23}
    H=\pm\left(\frac{8\pi^2}{\kappa\hbar N}\right)^{1/2}.
\eeq
  It follows from (\ref{mar-eq21}), (\ref{mar-eq22}) that on the imaginary
  axis $\eta=0, \upsilon\neq 0$ one gets (\ref{mar-eq17}), and on the real
  axis $\eta\neq 0, \upsilon=0$, one gets (\ref{mar-eq6}). Thus we arrive
  at the identity
\beq                \label{mar-eq24}
    {H_\tau}^2 \upsilon^2 \equiv H^2 \eta^2
\eeq

  At the origin $\varsigma = 0+i0$, we choose the second solution of
  (\ref{mar-eq20}), $H=H_{\tau}=0$, which provides the junction condition
  $(a'/a^2)_{\varsigma=0}=0$ at the boundary of signature change (see [22,
  23]). In the same figurative words, the analytic continuation to real
  time ``de-materializes'' the ghosts, and in real Lorentzian space they
  become fictitious particles.\footnote
    {This fact distinguishes the interpretation of the de Sitter exact
    solution in this paper from [32], where ``ghost materialization''
    takes place in real time.}
  The numerical value of $N$ is of the order of the number of gravitons
  tunneled into de Sitter space of real time by instantons (see below).
  In the case of the Poisson distributions (used above), $N^{-1}\sim
  \aver{ (\Delta N/N)^2 }$, where $\Delta N$ is a fluctuation of the number
  of gravitons. This means that $H^2\sim \aver{ (\Delta N/N)^2 }$, so that
  the speed of Hubble expansion is governed by quantum metric fluctuations,
  as expected. From (\ref{mar-eq15}) follows the equation of state
\[
    \eps_g = -p_g = \frac{3\hbar N}{8\pi^2}H^4.
\]

  This equation of state is superficially similar to what comes from quantum
  conformal anomalies. As was shown by Starobinsky [36], quantum corrections
  to the Einstein equations due to zero oscillations can provide a
  self-consistent de Sitter solution in the vicinity of Planck's value of
  curvature (see also [37]). In such a case, the equation of state is $\eps
  \sim n \hbar H^4$ [38], where the number of types of elementary particles
  $n$ is $\leq 100$. Conformal anomalies that arise due to regularization
  and renormalization procedures do not apply to this work, which deals with
  the equations of quantum gravity that are finite in the one-loop
  approximation. In the finite one-loop quantum gravity, the effect of
  conformal anomalies is exactly zero, and the de Sitter solution can be
  formed only by graviton-ghost instantons ([32], Sec. XII). In contrast
  to the conformal anomaly parameters, the parameter $N$ is arbitrary and
  can be a huge number.

  Using a mathematical analogy between (\ref{mar-eq4}), (\ref{mar-eq5}),
  (\ref{mar-eq11}), (\ref{mar-eq12}) and the stationary Schr\"{o}dinger
  equation, solutions to these can be thought of in terms of quantum
  tunneling. In these equations, $x = k\eta$ plays the role of the spatial
  coordinate of the Schr\"{o}dinger equation, and the role of a
  ``one-dimen\-sional potential'' is played by $a''/a$.  Whether
  (\ref{mar-eq4})--(\ref{mar-eq5}) belong to Lorentzian and
  (\ref{mar-eq11})--(\ref{mar-eq12}) to Euclidean space is governed by
  the sign of $k^2$ (where $+k^2$ is for real and $-k^2$ is for imaginary
  time). Superhorizon gravitons and ghosts $(|x^2| \ll 1)$ do not ``feel''
  the difference between Lorentzian and Euclidean signatures and can belong
  to each of these. This means that the boundary $x=0$ plays the role of a
  classically impenetrable barrier dividing these topologically
  non-equivalent spaces. To complete the analytic continuation of the
  self-consistent solution (\ref{mar-eq13}) and (\ref{mar-eq17}) from
  imaginary to real time, we should do that for the graviton and ghost mode
  functions. To do so, we analytically continue $g(\xi)$ of (\ref{mar-eq14})
  from the imaginary axis $\upsilon$ to the plane of complex conformal time
  $\varsigma=i\eta+\upsilon$. It reads
\beq                                   \label{mar-eq25}
    g(\zeta) = \left(1+\frac{1}{\zeta}\right) e^{-\zeta},
    \quad  \zeta = k\varsigma = ix + \xi.
\eeq

  It follows from (\ref{mar-eq25}) that on the imaginary axis $x=0,\ \xi\neq
  0$ $g(\zeta)=g(\xi)$ in accordance with (\ref{mar-eq14}), while on the
  real axis $x\neq 0,\ \xi=0$, one gets $g(\zeta)=f(x)$ from (\ref{mar-eq8}).
  The latter is an analytical continuation of $g(\xi)$ from (\ref{mar-eq14})
  to the space of real time. Thus, in the space of real time one gets a
  self-consistent solution consisting of (\ref{mar-eq6}) and (\ref{mar-eq7}).
  The junction conditions on the boundary of signature change are
  $\dot{\hpsi}_{{\bf k}\sigma} = \dot{\hat{\vartheta}}_{{\bf k}} = 0$ [23].
  At the origin $\zeta=i0+0$, they are satisfied automatically for the mode
  functions $g/a$ and $f/a$.

  To decide whether the de Sitter gravitational instanton is physically
  allowed, one needs to calculate the Euclidean action $S_E$ to make sure
  that it is finite. The action $S_E$, as defined in 4D space with a
  positive signature, reads ([32], Sec. VII.A.2)
\bearr                                            \label{mar-eq26}
    S_E =\frac{1}{\kappa} \int d\tau \biggl\{
    3\left[ a^2\frac{d^2a}{d\tau^2} + a\left(\frac{da}{d\tau}\right)^2\right]
\nnn\ \
    + \frac{1}{8} \sum_{{\bf k}\sigma} \left(
    a^3 \frac{d\hpsi_{{\rm {\bf k}}\sigma}^{+}}{d\tau}
    \frac{d\hpsi_{{\bf k}\sigma}}{d\tau}
    + ak^2\hpsi_{{\bf k}\sigma}^+ \hpsi_{{\bf k}\sigma}\right)
\nnn \ \
    - \frac 14 \sum_{\bf k}\,\left(a^3 \frac{d\hat{\theta}_{\bf k}^+}
    {d\tau} \frac{d\hat{\theta}_{\bf k}}{d\tau}
    +ak^2\hat{\theta}_{\bf k}^{+}\hat{\theta}_{\bf k}\right) \biggr\}.
\ear

  The substitution of $a, \hpsi_{{\bf k}\sigma}$ and $\hat{\vartheta}_{\bf k}$
  from (\ref{mar-eq13}) and (\ref{mar-eq14}) to (\ref{mar-eq26}) transforms
  the integrand to the left-hand side of \eq (\ref{mar-eq19}), which is
  identically zero. The finiteness of Euclidean action justifies the fact
  that the self-consistent solution (\ref{mar-eq13}), (\ref{mar-eq14}) and
  (\ref{mar-eq17}) can be thought of as a gravitational instanton. The
  tunneling probability is proportional to $\exp(-S_E)$, which is unity in
  this case and means that the macroscopic evolution of the Universe is
  determined. The existence of the analytic continuation and finiteness of
  $S_E$ allow for proposing the birth of a flat inflationary Universe by
  tunneling from ``nothing'' by means of a de Sitter gravitational instanton
  built up by quantum metric fluctuations.\footnote
    {The present lack of a consistent quantum theory of gravity leads to
    the fact that the appearance of self-consistent de Sitter space in
    real time by tunneling admits ambiguous physical interpretations.
    According to [32], Sec. VII.2 and VII.3.B, the same solution (17)
    can be obtained by a procedure borrowed from Quantum Chromodynamics
    which corresponds to a different type of tunneling.}
  Note that the self-consistent solution obtained is valid in the
  applicability of the one-loop approximation because it is formed by
  over-horizon metric fluctuations. Therefore, it is likely to be regarded
  as a qualitative result showing a possible path of the cosmological
  evolution of the Universe. The analysis of further evolution of a new
  inflating Universe born in such a way is the subject of inflation theory
  (see, e.g., [39] for references to original works and [40] for an analysis
  of current problems).  It is useful to recall here that ``so far, the
  details of inflation are unknown, and the whole idea of inflation remains
  a speculation, though one that is increasingly plausible'' ([41], p.\,202).
  Recall that the Hubble constant $H$ is determined by the number of
  gravitons $N$ that were tunneled into de Sitter space by instantons
  carrying out the tunneling. In the case of DE birth (Section 4), for the
  observed value of the Hubble constant $H = 73.8\pm 2.4\ {\rm km\cdot
  s^{-1}\cdot Mpc^{-1}}$ [9], from (\ref{mar-eq24}) one gets $N\sim 10^{122}$
  (a numerical coefficient of the order of unity is omitted). If the
  cosmological constant $\Lambda$ is due to the vacuum energy, as generally
  accepted, then $L^2_p/\Lambda^{-1}\approx 10^{-122}$, where
  $L_p = \sqrt{G\hbar/c^3}$ is the Planck length, and it is in direct
  contradiction to observations. Although there are several attempts to avoid
  this problem (see, e.g., [42]), it still exist. As follows from
  (\ref{mar-eq23}), an alternative interpretation of this huge number is
  that it is simply the number of gravitons that have tunneled into the
  contemporary Universe, and it has nothing to do with the vacuum energy.

\section{Birth of dark energy}

  It follows from the observational data that the Universe consists of
  approximately $70\%$ of DE and $30\%$ of dark matter plus ordinary matter
  in the present epoch. In the previous section, we showed that tunneling
  from ``nothing'' may create the de Sitter expansion of the empty Universe
  in real time. In this section, we consider this process in the presence of
  matter. The aim of this section is to address the ``coincidence problem''
  and the specific features of the matter-dominated epoch that distinguishes
  it from others.

  In such a case, \eq (\ref{mar-eq2}) reads
\bearr                                     \label{mar-eq27}
    3\frac{{a'^2}_\eta}{a^4}=\kappa(\eps_g+\eps_m),
    \quad  \kappa\eps_m = \frac{3C_m}{a^3},
\nnn
    3C_m = \kappa\eps_m a^3 = \kappa\eps_{0m}{a_0}^3 = \const.
\ear
  Here $\eps_{0m}$ and $a_0$ are the present energy density of matter and
  scale factor, respectively, and $\eps_g$ is taken from (\ref{mar-eq2}).
  In accordance with the general idea of this work, one must obtain a
  solution to the set of equations (\ref{mar-eq27}), (\ref{mar-eq4}) and
  (\ref{mar-eq15}) in imaginary time and then analytically continue it
  to the Lorentzian space of real time. In imaginary conformal time
  $\varsigma$, (\ref{mar-eq27}) reads
\beq                                  \label{mar-eq28}
    3\frac{a'^2}{a^4}=-\kappa (\eps_{\rm inst}+\eps_m).
\eeq

  The energy density of graviton-ghost instantons $\kappa\eps_{\rm inst} $
  is here the right-hand side of \eq (\ref{mar-eq10}). Primes in this
  section denote derivatives in the imaginary conformal time $\upsilon$.
  Solutions to \eq (\ref{mar-eq28}) can exist only if $\eps_{\rm inst}<0$,
  i.e., again under the condition that ghosts are ``materialized'' in
  Euclidean space. This condition is necessary but not sufficient. In the
  presence of matter, solutions to (\ref{mar-eq28}) can exist only after the
  scale factor $a(\varsigma)$ exceeds the threshold $a\geq a_{\rm treshold}
  \equiv a_{T}$, which is determined by the condition $\eps_m \leq
  -\eps_{\rm inst}$. In the framework of our assumption that the DE is of
  instanton origin and that the imaginary-time solution can be analytically
  continued to real time, this threshold condition is transformed to
  $\eps_m \leq \eps_{\rm de} $. As was mentioned in Section 1, the
  observational data are consistent with the de Sitter expansion law, so
  that $\eps_{\rm de}\approx \const$.\footnote
    {In the general case, if $\eps_{\rm de} =\eps_{\rm de} (t)\ne
    \const$, the threshold condition can be redefined as $\eps_{m} \le
    \eps_{\rm de}^{(\min)}$, where $\eps_{\rm de}^{(\min )}$ is the
    deepest minimum of $\eps_{\rm de}$ (see the figure).}
  From this fact and (\ref{mar-eq27}) it follows that for
  $z_{\rm theshold}\equiv z_T$ one gets
\beq                                           \label{mar-eq29}
    1 + z_{T} \leq \left(\frac{\eps_{\rm de}}{\eps_m}\right)^{1/3}
    = \left( \frac{\Omega_{\rm de}}{\Omega_m} \right)^{1/3}.
\eeq

  The last term in (\ref{mar-eq29}) is presented with the generally accepted
  notation where $\Omega_{\rm de}$ and $\Omega_m$ are the ratios of DE
  density and matter density to the total density of the Universe,
  respectively, so that $\Omega_{\rm de}+{\Omega_m}=1$. Assuming that
  $\Omega_{\rm de}\approx 0.69$ [43], one gets $1 + z_T \leq 1.3$.
  In the existence of such a threshold lies a possible answer to the
  question: why the birth of DE occurred ``recently'', i.e. after its energy
  density became comparable with the energy density of matter $\eps_m$?

  The decelerated expansion changed to an accelerated one in real time at
  the transition point $z_t$  where $\ddot{a}=0$. According to [44],
  $z_t = 0.35\pm 0.07$; by [45] $z_t \approx 0.29_{-0.06}^{+0.07}$ to
  $z_t\approx 0.60_{-0.08}^{+0.06}$; by [46], $z_t\sim 0.3$; by [47],
  $z_t\approx 0.43\pm 0.07$; by [32], Sec. IX, $z_t \approx 0.34\pm 0.02$;
  by [48], $z_t \approx 0.78^{+0.08}_{-0.27}$. Blake et al. [49] and
  Busca et al. [50] did not estimate $z_t$ but found that an acceleration
  can take place for $z < 0.7$. In the case of a $\Lambda$ term and the
  same assumption $\Omega_{\rm de} \approx 0.69$, the condition $\ddot{a}=0$
  leads to the transition point redshift
\beq                                      \label{mar-eq30}
    z_t^{(\Lambda)} =\biggl(\frac{2\eps_{\Lambda}}{\eps_m}\biggr)^{1/3}-1
    = \biggr(\frac{2\Omega_{\Lambda}}{\Omega_m}\biggr)^{1/3}-1 = 0.64.
\eeq

  The presence of large systematic and statistical errors makes it difficult
  to distinguish between $z_t$ that come from the observational data listed
  above and $z_t^{(\Lambda)}$ that comes from a $\Lambda$-term theory. We
  would like to emphasize once again that although both the cosmological
  constant and instanton DE asymptotically approach the same de Sitter
  regime where the impact of matter becomes negligible, but this does not
  mean that in the transition region, where the DE and matter densities are
  comparable, transition points must be the same. Most likely they should
  differ because \eqs (\ref{mar-eq27}) are different for DE of instanton
  origin and a cosmological constant. To see that, one can combine \eqs
  (\ref{mar-eq2}) and (\ref{mar-eq3}) and pass over from the conformal real
  time to the physical imaginary time $\tau$. As a result, one gets
\beq                                                   \label{mar-eq31}
    \frac{\ddot{a}_{\tau}}{a}=\frac{\kappa}{6}
        (\eps_m+\eps_{\rm inst} + 3p_{\rm inst}).
\eeq
  If $p_{\rm inst}\to 0$ in the vicinity of the threshold point, then the
  difference between $z_t$ and $z_T$ also tends to zero, and this is often
  the case, as shown by numerical experiments (see the figure). Unlike the
  cosmological constant case, where $z_t$ is about twice as large as $z_T$,
  they can be close to each other in the case of DE of instanton origin.

\begin{figure*}
\centering
\includegraphics*[width=0.65\textwidth]{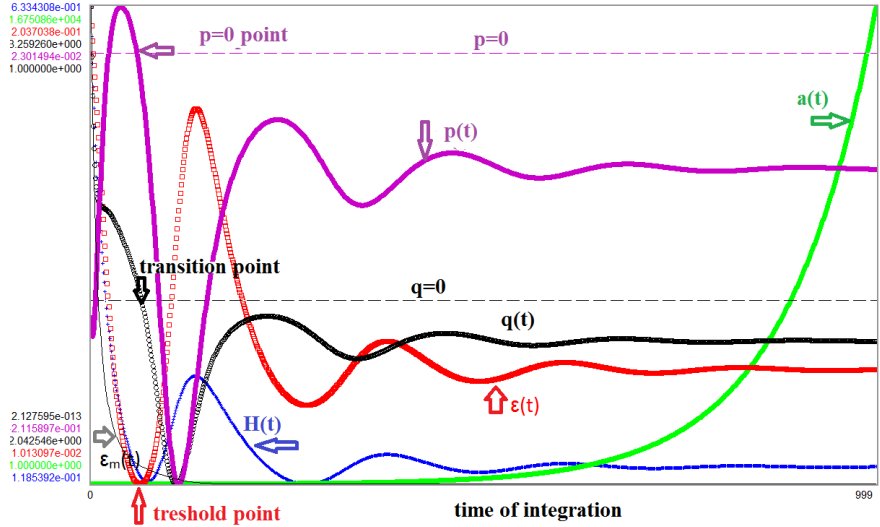}
\caption{\small
  A real-time numerical solution is shown for the cosmological substrate
  consisting of DE of instanton origin and cold dark matter. Here $a(t)$ is
  the scale factor; $H(t)=\dot{a}/a$ is the Hubble function;
  $q(t)=-\ddot{a}/a$ is the acceleration/deceleration parameter;
  $\eps_m(t)$ is the energy density of matter; $\eps(t)$ and $p(t)$ are
  the energy density and pressure of DE. All functions are dimensionless but
  plotted in different scales for the sake of clarity. Numbers along the
  Y axis indicate numerical values of maximal (upper group) and minimal
  (lower group) values of the appropriate functions. The upper and lower
  dashed lines correspond to the zeros of the functions $p(t)$ and $q(t)$,
  respectively. At the transition point, the scale factor is $a_t=2.99$. Here
  $q(t)$ passes through zero, changes its sign, and it is shifting away from
  deceleration $(q(t)>0)$ to acceleration $(q(t)<0)$. In the vicinity of
  this point, the pressure $p$ passes through zero at the scale factor
  $a_p=2.84$, and $\eps$ approaches its deepest minimum at the threshold
  point (the scale factor $a_T=2.96$ ). For this particular example, one
  gets $(a_T-a_t)/a_t\approx 0.01$ and $(a_{p=0}-a_t)/a_t\approx 0.05$. So,
  $z_t - z_T = (\Delta a/a)(1+z_t) \approx 0.01 (1+z_t)$, and
  $z_{p=0} - z_t \approx 0.05(1+z_t)$, respectively. All these differences
  lie within the range of observational errors. As can be seen from this
  graph, the solution oscillates from the start. The nature of these
  oscillations (the number of maxima and minima, their amplitudes, etc.)
  depends on the choice of initial conditions which are unknown.
  Asymptotically, however, all such solutions come to the de Sitter mode
  $H\to const$; $-q\to H^2$; $\eps\to 3H^2$; $p\to -3H^2$. This fact is
  independent of specific initial conditions if they are of the type that
  corresponds to the de Sitter attractor.  }
\label{mar-fig1}
\end{figure*}

  Unlike Section 3, where we had an exact and explicit de Sitter solution,
  we have no explicit solutions to the set of equations (\ref{mar-eq11}),
  (\ref{mar-eq12}) and (\ref{mar-eq28}) here to analytically continue them
  to the space of real time. To get numerical solutions in real time, we can
  employ the BBGKY-chain built on \eqs (\ref{mar-eq27}), (\ref{mar-eq4}),
  (\ref{mar-eq5}). In an explicit form, they are \eqs (VIII.2)--(VIII.4)
  of [32]. The output is a numerical solution in real time. The figure shows
  a typical real-time numerical solution by means of a BBGK-chain for rather
  arbitrary initial conditions.\footnote
    {As was shown in [32], section V, the BBGK chain built on \eqs
    (\ref{mar-eq2}), (4) and (5) has at least three attractors, and the
    de Sitter instanton is just one of them.  Accordingly, the
    asymptotic output on this or another attractor is determined by the
    choice of initial conditions. Therefore, in the case shown in the
    figure, the initial conditions are rather arbitrary but within those
    leading to the de Sitter attractor.}
  At the beginning, oscillations can be seen, whose nature depends on the
  initial conditions. Then, oscillations decay, and the solution
  asymptotically approaches a de Sitter expansion mode when the influence of
  matter becomes negligible. The exit to the de Sitter regime with decreasing
  matter input, regardless of the specific initial conditions (if the
  latter correspond to the de Sitter attractor) is not a specific property
  of the present numerical example. It is a general property of the theory
  of DE of instanton origin. To show that, let us consider an approximate
  but explicit solution to \eq (\ref{mar-eq28}). It can be obtained in the
  asymptotic case $|\eps| \gg \eps_m$ when the energy density of matter is
  small as compared to that of instantons. In such a case, one can replace
  $-\kappa \eps_{\rm inst}\approx 3 H^2_{\tau} = \const$ in \eq
  (\ref{mar-eq28}), where $H^2_{\tau}$ is defined by (\ref{mar-eq20}).

  Such a replacement, meaning closeness of the initial conditions to the
  de Sitter attractor, leads to \eq (\ref{mar-eq32}) instead of
  (\ref{mar-eq28}), which can be solved explicitly:
\beq                \label{mar-eq32}
    3\frac{\dot{a^2_{\tau}}}{a^2}=3H^2_{\tau}-\frac{3C_m}{a^3}
\eeq
  The solution to (\ref{mar-eq32}) reads
\bearr                                     \label{mar-eq33}
    a(\tau) = a_0 \biggl[\cosh\biggl(\frac{3}2H_{\tau}\tau\biggr)
\nnn\cm
    +\sqrt{1-\frac{C_m}{a^3_0 H^2_{\tau}}}
    \sinh\biggl(\frac{3}2H_{\tau}\tau\biggr)\biggr]^{2/3}.
\ear

  In the absence of matter $(C_m=0)$, the solution (\ref{mar-eq33})
  transforms into (\ref{mar-eq17}), i.e., pure de Sitter expansion. In the
  presence of matter $(C_m\neq 0 )$, the solution (\ref{mar-eq33}) clearly
  demonstrates the existence of the threshold $C_m/(a^3_0 H^2_{\tau})
  =\eps_{0,m}/|\eps_{\rm inst}|\leq 1$. A transition to real time in
  (\ref{mar-eq33}), taking into account (\ref{mar-eq9}) and
  (\ref{mar-eq22}), leads to
\bearr                        \label{mar-eq34}
    a(t)=a_0 \biggl[\cosh \Big(\frac{3}2Ht\Big)
\nnn \cm    
	+\sqrt{1+\frac{C_m}{a^3_0 H^2}}
    \sinh \Big(\frac{3}2Ht\Big)\biggr]^{2/3} .
\ear
  Indeed, it coincides with the well-known solution for a cosmological
  model containing a positive cosmological constant in the presence of
  non-relativistic matter. In the latter case, this solution is valid for any
  instant of time. Unlike that, in our case, the \eq (\ref{mar-eq32})
  itself, its solution (\ref{mar-eq33}) and the real time solution
  (\ref{mar-eq34}) are valid only asymptotically when the contribution of
  matter is small. Asymptotically, for $3H_{\tau}\tau/2 \gg 1$ and
  $3Ht/2 \gg 1$, one gets from (\ref{mar-eq33}), (\ref{mar-eq34})
\bearr                   \label{mar-eq35}
    a(\tau)=a_0 e^{H_{\tau}\tau} \biggl[\frac 12
    \biggl(1+\sqrt{1-\frac{C_m}{a^3_0 H^2_{\tau}}}\biggr)\biggr]^{2/3},
\yyy            \label{mar-eq36}
    a(t)=a_0 e^{Ht} \biggl[\frac 12 \biggl(1 +
            \sqrt{1+\frac{C_m}{a^3_0 H^2}}\biggr)\biggr]^{2/3}.
\ear

  In conformal time, (\ref{mar-eq35}) and (\ref{mar-eq36}) become
  (\ref{mar-eq17}) and (\ref{mar-eq6}), respectively, and the identity
  (\ref{mar-eq24}) provides again a possibility of analytical continuation
  of (\ref{mar-eq17}) to (\ref{mar-eq6}) over the horizon $\eta=0$. Thus,
  with increasing of the scale factor, its behavior is getting
  asymptotically closer and closer to the de Sitter regime. The smaller the
  contribution of matter to the energy balance, the more conditions for
  tunneling approach those for empty space, and the latter are satisfied
  exactly in accordance with section 3. Apart from having a threshold and
  the de Sitter asymptotic  behavior, there is another specific feature of
  the matter-dominated era, which distinguishes it from other eras. It is
  that DE tunneling to the real-time Universe has favorable conditions
  precisely in this era (see below).

  To complete the analytic continuation of the self-consistent solution, we
  must carry the graviton and ghost mode functions across the barrier $x=0$
  in the presence of matter. As was already mentioned, using a mathematical
  analogy between (4), (5), (\ref{mar-eq11}), (\ref{mar-eq12}) and
  the stationary Schr\"odinger equation, solutions to these can be thought
  of in terms of quantum tunneling. For $a = \const\cdot\eta^{-\beta}$, the
  ``one-dimensional potential'' is $a''/a = \beta (\beta+1)/\eta^2$ [51].
  A remarkable fact is that the ``one-dimensional potentials''
  $a''/a = 2/\eta^2$ are the same in both cases $\beta=-2$ (a matter
  dominated background with the equation of state $p=0$) and $\beta=1$
  (a de Sitter background with the equation of state $p=-\eps$). The same is
  also true for the imaginary time $\varsigma$. Since the ``one-dimensional
  potentials'' coincide, the Schr\"odinger-like equations (4), (5)
  for gravitons and ghosts over the matter dominated background and over the
  de Sitter background are identical. The same is true for \eqs (11), (12).
  Due to identity of equations for matter-dominated and de Sitter
  backgrounds, the boundary conditions for tunneling are naturally satisfied
  at the barrier. It can be seen from two opposite limiting cases
  $\eps_m \ll |\eps_{\rm inst}|$ and $\eps_m \gg |\eps_{\rm inst}|$.

  As was shown in Section 3, in the first limiting case, the boundary
  conditions are satisfied by the functions $f(x)$ from (7), (8),
  and $g(\xi)$ from (\ref{mar-eq14}), which correspond to the solutions of
  equations (4), (5), (\ref{mar-eq11}) and (\ref{mar-eq12}) in the
  de Sitter background, $a\sim \eta^{-1}$ in real time and $a\sim
  \upsilon^{-1}$ in imaginary time. In the second limiting case, one gets
  the same function $f(x)$ in the matter-dominated background $a\sim
  \eta^2$ and $g(\xi)$ in the de Sitter background $a\sim \upsilon^{-1}$.
  In the case of interest (the first limiting case), we can consider
  (\ref{mar-eq17}) and (\ref{mar-eq25}) as approximations valid for scale
  factors that are close to de Sitter ones in both imaginary and real
  times. The combination of the two facts, the existence of a threshold and
  coincidence of ``one-dimensional potentials'', distinguishes the
  matter-dominated epoch from others. As a result, we arrive at the following
  picture. With a decreasing contribution of matter, the Universe is
  increasingly emptied, and conditions for tunneling approach those for
  empty space. Because of the identity (\ref{mar-eq24}), nothing prevents an
  empty Universe from tunneling back to ``nothing'' at the end of its
  cosmological evolution.

\section{Conclusion}

  In imaginary time, quantum metric fluctuations of empty Euclidean space
  form an exact solution to the self-consistent equations of quantum gravity
  in the one-loop approximation that can be thought of as a de Sitter
  gravitational instanton. This solution is analytically continued into the
  Lorentzian space of real time where it gives rise to a de Sitter
  expansion. In the presence of matter, the same effect is switched on
  after the energy density of matter drops below a threshold. The following
  scenario can be proposed. A flat inflationary Universe could have been
  formed by tunneling from ``nothing''. After that it should evolve
  according to inflation scenarios that are beyond the scope of this paper.
  Then the standard Big Bang cosmology starts and lasts as long as the
  Universe begins to become empty again. As the Universe ages and is
  emptied, the same mechanism of tunneling that gave rise to the empty
  Universe at the beginning, gives now birth to dark energy. This mechanism
  is switched on after the energy density of matter has dropped below a
  critical level. After that, to the extent that the space continues to be
  empty, the expansion proceeds faster and faster and gradually becomes
  again exponentially fast (de Sitter). The identity (\ref{mar-eq24})
  provides a possibility for the empty Universe (that has completed its
  cosmological evolution) to be able to tunnel back to ``nothing''. After
  that, the entire scenario can be repeated indefinitely.


\subsection*{Acknowledgment}

  Key elements of the proposed scenario are the Faddeev-Popov ghosts, without
  which instantons cannot create a self-consistent de Sitter solution in
  imaginary time. Therefore, the mathematical correctness of the basic
  equations of quantum gravity, which suggests inevitability of the
  appearance of ghosts in the theory of quantum metric fluctuations, has
  been the subject of my special attention. I would like to express my deep
  appreciation to Ludwig D. Faddeev of the Steklov Mathematical Institute for
  graciously agreeing to read our manuscript by Vereshkov and Marochnik [33]
  on the role of Faddeev-Popov ghosts, equations of quantum gravity in the
  Heisenberg representation, and verifying the correctness of our approach.
  I am deeply grateful to Mikhail Shifman and Arkady Vainshtein of the
  University of Minnesota for discussions of the structure and content of the
  theory presented in the manuscript noted above. My deepest gratitude to
  Grigory Vereshkov of South Federal University (Russia) for numerous
  discussions on the key points raised in this paper that include but are not
  limited to Faddeev-Popov ghosts, instantons, imaginary time and fictitious
  fields of inertia in quantum gravity. I am deeply grateful to Roald
  Sagdeev and Daniel Usikov of the University of Maryland, Boris Vayner of
  NASA Glenn Space Center and Arthur Chernin of Sternberg Astronomical
  Institute for useful discussions. I am grateful to Leonid Grishchuk of
  Cardiff University who sent the preprint by Grishchuk and Zeldovich [17]
  at my request and made useful comments on this work. I am grateful to
  Yuri Shchekinov and Victoria Yankelevich for help in the manuscript
  preparation. Also, I would like to express my deep appreciation and special
  thanks to my friend and colleague Walter Sadowski for invaluable advice and
  help in the preparation of the manuscript. I am grateful to the anonymous
  referee for valuable comments that helped to improve this paper.

} 

\end{document}